# Large Negative Magnetoresistance in off-Stochiometric Topological Material PrSbTe

Gokul Acharya[1], Krishna Pandey[2], M.M. Sharma[1], Jian Wang[3], Santosh Karki Chhetri[1], Md Rafique Un Nabi[1], Dinesh Upreti[1], Rabindra Basnet[4], Josh Sakon[5], Jin Hu[1,2*]

[1]Department of Physics, University of Arkansas, Fayetteville, AR, USA

[2]Materials Science and Engineering Program, Institute for Nanoscience and Engineering, University of Arkansas, Fayetteville, AR, USA

[4]Department of Chemistry and Biochemistry, Wichita State University, Wichita, KS, USA

[4]Department of Chemistry & Physics, University of Arkansas at Pine Bluff, Pine Bluff, Arkansas 71603, USA

[5]Department of Chemistry and Biochemistry, University of Arkansas, Fayetteville, Arkansas 72701, USA

## Abstract

Magnetic topological materials *Ln*SbTe (*Ln* = lanthanide) have attracted intensive attention because of the presence of interplay between magnetism, topological, and electron correlations depending on the choices of magnetic *Ln* elements. Varying Sb and Te composition is an efficient approach to control structural, magnetic, and electronic properties. Here we report the composition-dependent properties in $PrSb_xTe_{2-x}$. We identified the tetragonal-to-orthorhombic structure transitions in this material system, and very large negative magnetoresistance in the $x$ =

0.3 composition, which might be ascribed to the coupling between magnetism and transport. Such unusual magnetotransport enables $PrSb_xTe_{2-x}$ topological materials as a promising platform for device applications.

*jinhu@uark.edu

## I. INTRODUCTION

An extensively studied phenomenon in condensed matter physics is the response of magnetic materials under external magnetic field. Large magnetoresistance characterized by strong modulation of resistivity under magnetic field, such as giant magnetoresistance (GMR) and colossal magnetoresistance (CMR) have potential applications for technological advances like spintronic devices and magneto resistive read heads in the magnetic recording industry [1,2]. GMR in magnetic multilayers is caused by reduced spin scattering when magnetic moments are aligned parallelly by magnetic field [1,3]. The mechanism for CMR is more complicated. CMR in manganate perovskite materials has been ascribed to ferromagnetic double exchange mechanism between mixed valence of $Mn^{3+}$/$Mn^{4+}$ and a structural Jahn-Teller distortion [4–8].

The discovery of topological semimetal materials (TSMs) in the past two decades has greatly enriched the magnetotransport phenomenon. Topological semimetals (TSMs) such as Dirac and Weyl semimetals possess electronic band structures with linear band crossings protected by various symmetries, which host electrons with low-energy excitations resemble to relativistic Dirac or Weyl fermions [9–12]. Upon applicating magnetic field, large positive magnetoresistance characterized by extremely strong resistivity increase has been discovered in many non-magnetic TSMs such as $Cd_3As_2$ [13], $PtBi_2$ [14], $WTe_2$ [15], and NbP [16]. In addition, TSMs also display exotic chiral anomaly in magnetotransport, which is manifested as negative magnetoresistance when magnetic field is aligned along the electric field (i.e., electric current) direction [17–21].

In addition to non-magnetic TSMs, topological states have also been discovered in various magnetic materials such as $HgCr_2Se_4$ [22], $Co_3Sn_2S_2$ [23–25], Heusler compounds [26–28], and $Mn_3$(Sn/Ge) [29–33]. Besides these materials in which magnetism originates from 3*d* transition metal, the study has been extended to 4*f*-magnetism in rare earth compounds such as *Ln*SbTe [34–46] (*Ln* = lanthanide) and GdPS [47], in which the magnetism is brought in by 4*f* rare earth elements. Structurally, these materials can be viewed as deviations of non-magnetic ZrSiS, which crystallizes in a tetragonal PbFCl-type structure (space group *P*4/*nmm*) with two-dimensional (2D) square nets of Si sandwiched by Zr-S layers [48,49]. Unlike Dirac or Weyl semimetals that possess discrete Dirac or Weyl points in the momentum space, ZrSiS displays linear band crossings along a line in its electronic band structure and is thus classified as topological nodal-line semimetal. In addition to many non-magnetic variation of ZrSiS via replacing Zr, Si, and S elements [48–55], spin degree of freedom is activated in magnetic *Ln*SbTe and GdPS and leads to exotic properties. In *Ln*SbTe topological materials, AFM ground state has been identified for various *Ln* = Ce [36,56,57], Nd [34,58,59], Sm [39,60,61], Gd [35,62,63], Tb [42,43], Dy [44,45], Ho [38,43,46], and Er [45], which results in metamagnetic transitions [34,36–39,62], magnetic devil's staircase [37,57], and possible electronic correlation enhancement [38,39]. Magnetism is also predicted to modulate time-reversal symmetry and creates topological phase transition [36]. In GdPS, the strong *d*-*f* magnetic exchange interaction of Gd alter electronic structure and leads to an insulator-to-metal transition [47]. Furthermore, the isotropic magnetism owing to minimized magnetic anisotropy of $Gd^{3+}$ causes distinct isotropic gigantic magnetoresistance [47].

One unique property of *Ln*SbTe is composition non-stoichiometry that enables additional tuning. In these materials, Dirac fermions are generated by the Sb plane. With substituting Sb by Te to form $Ln\mathrm{Sb}_x\mathrm{Te}_{2-x}$, the lattice evolves from tetragonal (space group *P4/nmm*) for stoichiometric and nearly stoichiometric *Ln*SbTe, to orthorhombic for strongly off-stoichiometric compositions [57,60,63–65]. For some $Ln\mathrm{Sb}_x\mathrm{Te}_{2-x}$ such as $\mathrm{NdSb}_x\mathrm{Te}_{2-x}$, tetragonal phase re-emerges when Sb content is very low [64]. The tetragonal-to-orthorhombic distortion is accompanied by charge density waves (CDWs) which interplays with topological states and magnetism [66]. Therefore, the composition stoichiometry in $Ln\mathrm{Sb}_x\mathrm{Te}_{2-x}$ provides another approach to engineer electronic states and properties. In this work, we focus on the less-explored, off-stoichiometric $\mathrm{PrSb}_x\mathrm{Te}_{2-x}$ [67]. The existence of Dirac nodal-line state has been identified in stoichiometric PrSbTe [40,41]. In off-stoichiometric compositions studied in this work, we found the tetragonal to orthorhombic structural phase transition with composition tuning. Interestingly, well-defined long-range AFM order, which is commonly seen in magnetic *Ln*SbTe compounds, is absent in $\mathrm{PrSb}_x\mathrm{Te}_{2-x}$. Nevertheless, with carefully controlling the composition to $x = 0.3$, unusual strong negative magnetoresistance is observed. With such properties, $\mathrm{PrSb}_{0.3}\mathrm{Te}_{1.7}$ provides a new playground to investigate exotic magnetotransport without long-range magnetic orders.

## II. EXPERIMENT

Single crystals of $\mathrm{PrSb}_x\mathrm{Te}_{2-x}$ ($0.05 < x < 0.91$) were grown by a chemical vapor transport method with $I_2$ as the transport agent. Elementary powder of Pr, Sb and Te as the source materials were loaded in quartz tubes for chemical vapor transport, which was performed with a temperature gradient from 1000 to 850 °C for two weeks with source materials located at the hot end of the tube. Single crystals of various compositions were obtained by varying the ratio of Sb and Te in the source material, as shown in Fig. 1a. The compositions and crystal structures of the as-grown single crystals were examined by energy-dispersive x-ray spectroscopy (EDS) and x-ray diffraction (XRD). Electronic transport measurements were performed using the standard four-probe method in a physical property measurement system (PPMS, Quantum Design). Magnetization measurements were performed by using a magnetic property measurement system (MPMS3 SQUID, Quantum Design).

## III. RESULTS AND DISCUSSION

The nearly stoichiometric $x$ = 0.91 sample was examined by single crystal XRD refinement, which was found to display a tetragonal structure (space group P4/*nmm*) featuring Sb square nets. Such tetragonal structure is consistent with the reported structure for stoichiometric PrSbTe [40,41]. Upon Te substitution for Sb, samples with Sb content below 0.80 display an orthorhombic distortion with a space group of *Pmmn*, which is revealed by Rietveld refinement of powder XRD (Fig. 1a). Such tetragonal to orthorhombic structure evolution has also been widely seen in other *Ln*SbTe compounds [57,60,63–65]. Further increasing the Te content leads to a shrinkage of the $c$-axis and an expansion of the $a$ and $b$ axes. Interestingly, with sufficiently low Sb content ($x$ = 0.05), the tetragonal phase emerges again. The re-entering of tetragonal phase has not been observed in many *Ln*SbTe materials except for $NdSb_xTe_{2-x}$ [64]. Despite the structure evolution, the actual crystal structures for each composition are highly similar, as shown in Fig. 1b. The difference between tetragonal and orthorhombic structures can be seen from the evolution of the Sb-plane. As shown in Fig. 1c, the square net lattice for Sb in the tetragonal $x$ = 0.05 sample becomes distorted in the orthorhombic $x$ = 0.30 and 0.80 samples, as reflected by the deviation from 90° for the Sb-Sb bonding angles. The complete evolution of lattice parameters is summarized in Fig. 1d and the refined crystal structure parameters are provided in Table 1. Additionally, previous studies on some other *Ln*SbTe compounds have revealed the presence of vacancies in Sb layer in low-Sb compositions [57,60,63,68,69], which is not observed in our $PrSb_xTe_{2-x}$ within the instrumental resolution of EDS and XRD. As shown in Fig. 1d, for the orthorhombic compositions, the degree of orthorhombic distortion, characterized by $\frac{(a-b)}{(a+b)/2}$, is the strongest for the $x$ = 0.3 sample.

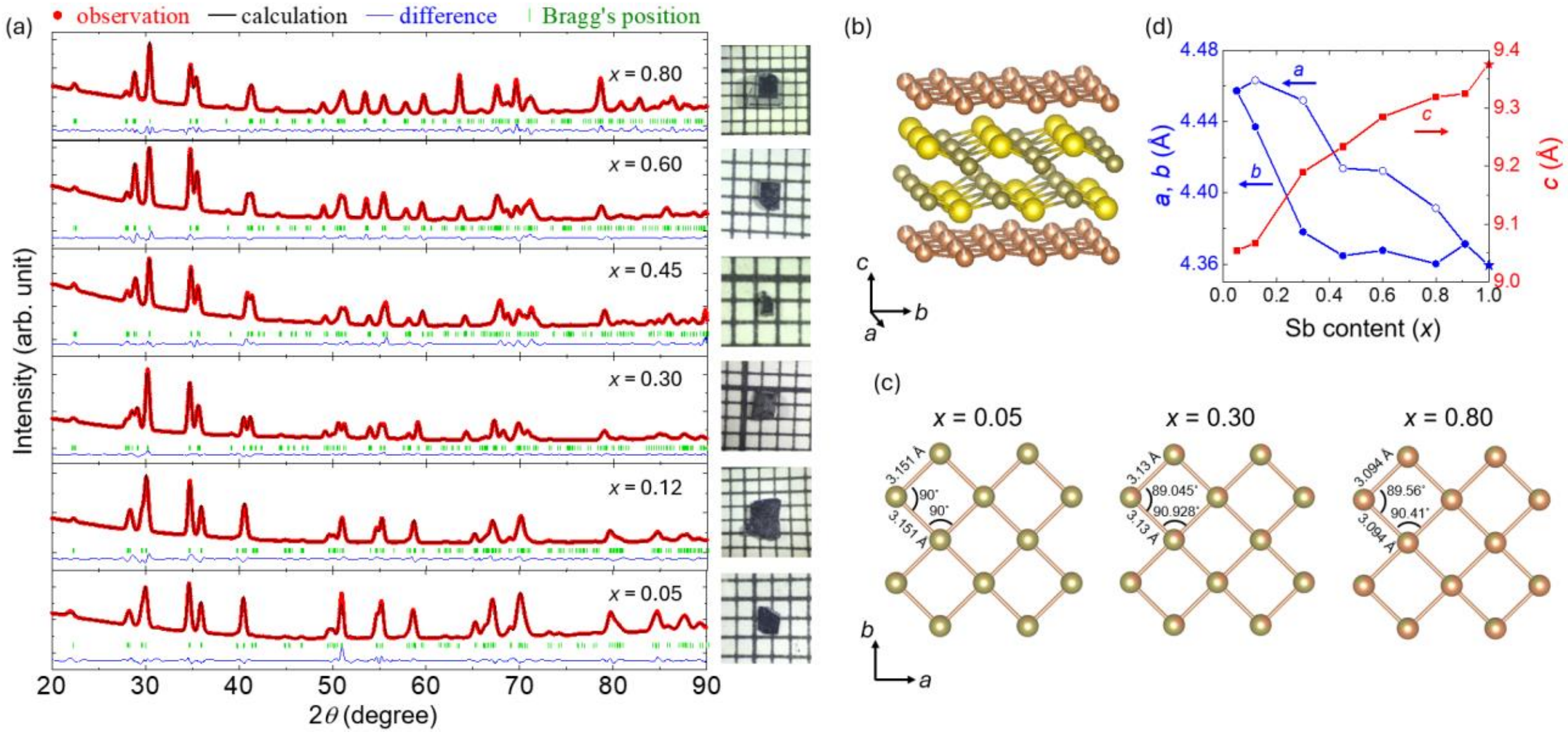

**Figure 1** (a) Rietveld refinement of the x-ray diffraction patterns for $PrSb_xTe_{2-x}$. The broadening and splitting of the peak near 40 degrees is an indication of orthorhombic distortion. The right panel shows the images of single crystals with various Sb contents. The square mesh measures 1 $mm^2$. (b) Crystal structure of $PrSb_xTe_{2-x}$. The tetragonal and orthorhombic structures are very similar. The difference between them can be seen in (c), the evolution of Sb plane with composition. (d) Evolution of lattice parameters $a$, $b$ and $c$ with varying Sb content. Data for x = 1 is taken from Ref. [41]

**Table 1**. Crystal structures for various $PrSb_xTe_{2-x}$ compositions obtained from powder XRD refinement, except for $PrSb_{0.91}Te_{1.09}$ which was obtained from single crystal XRD.

| | Space group | Lattice parameters (Å) | | | Atoms | Atomic positions | | | Occ. | *R*-factors |
|---|---|---|---|---|---|---|---|---|---|---|
| | | *a* | *b* | *c* | | *x* | *y* | *z* | | |
| $PrSb_{0.91}Te_{1.09}$ | *P*4/*nmm* | 4.371(2) | 4.371(2) | 9.325(1) | Pr | 0.2500 | 0.2500 | 0.3721 | 1 | $R_1$ = 0.0645 |
| | | | | | Sb | 0.7500 | 0.2500 | 0 | 0.91 | $WR_2$ = 0.1123 |
| | | | | | Te1 | 0.2500 | 0.2500 | 0.7244 | 1 | |
| | | | | | Te2 | 0.7500 | 0.2500 | 0 | 0.09 | |
| $PrSb_{0.8}Te_{1.2}$ | *Pmmn* | 4.403(4) | 4.363(1) | 9.323(8) | Pr | 0.2500 | 0.7500 | 0.2778 | 1 | $R_p$=3.54 |
| | | | | | Sb | 0.2500 | 0.2500 | -0.0090 | 0.8 | $R_{wp}$=7.33 |
| | | | | | Te1 | 0.2500 | 0.7500 | 0.6344 | 1 | $R_{Bragg}$=8.60 |
| | | | | | Te2 | 0.2500 | 0.2500 | -0.0090 | 0.2 | |
| $PrSb_{0.6}Te_{1.4}$ | *Pmmn* | 4.412(4) | 4.367(7) | 9.287(1) | Pr | 0.2500 | 0.7500 | 0.2733 | 1 | $R_p$=6.74 |
| | | | | | Sb | 0.2500 | 0.2500 | -0.0047 | 0.6 | $R_{wp}$=10.3 |
| | | | | | Te1 | 0.2500 | 0.7500 | 0.6306 | 1 | $R_{Bragg}$=13.1 |
| | | | | | Te2 | 0.2500 | 0.2500 | -0.0047 | 0.4 | |
| $PrSb_{0.45}Te_{1.55}$ | *Pmmn* | 4.414(3) | 4.364(5) | 9.233(4) | Pr | 0.2500 | 0.7500 | 0.2723 | 1 | $R_p$=1.65 |
| | | | | | Sb | 0.2500 | 0.2500 | -0.0052 | 0.45 | $R_{wp}$=2.94 |
| | | | | | Te1 | 0.2500 | 0.7500 | 0.6292 | 1 | $R_{Bragg}$=4.75 |
| | | | | | Te2 | 0.2500 | 0.2500 | -0.0052 | 0.55 | |
| $PrSb_{0.3}Te_{1.7}$ | *Pmmn* | 4.459(1) | 4.386(3) | 9.205(8) | Pr | 0.2500 | 0.7500 | 0.2740 | 1 | $R_p$=1.35 |
| | | | | | Sb | 0.2500 | 0.2500 | -0.0023 | 0.3 | $R_{wp}$=2.18 |
| | | | | | Te1 | 0.2500 | 0.7500 | 0.6280 | 1 | $R_{Bragg}$=3.57 |
| | | | | | Te2 | 0.2500 | 0.2500 | -0.0023 | 0.7 | |
| $PrSb_{0.12}Te_{1.85}$ | *Pmmn* | 4.463(1) | 4.437(8) | 9.065(1) | Pr | 0.2500 | 0.7500 | 0.2728 | 1 | $R_p$=2.08 |
| | | | | | Sb | 0.2500 | 0.2500 | -0.0009 | 0.12 | $R_{wp}$=3.51 |
| | | | | | Te1 | 0.2500 | 0.7500 | 0.6318 | 1 | $R_{Bragg}$=2.93 |
| | | | | | Te2 | 0.2500 | 0.2500 | -0.0009 | 0.85 | |
| $PrSb_{0.05}Te_{1.95}$ | *P*4/*nmm* | 4.45710 | 4.45710 | 9.05400 | Pr | 0.2500 | 0.7500 | 0.2745 | 1 | $R_p$=2.46 |
| | | | | | Sb | 0.2500 | 0.2500 | 0.0026 | 0.05 | $R_{wp}$=3.75 |
| | | | | | Te1 | 0.2500 | 0.7500 | 0.6290 | 1 | $R_{Bragg}$=10.7 |
| | | | | | Te2 | 0.2500 | 0.2500 | 0.0026 | 0.95 | |

Figure 2a displays temperature dependance of resistivity for various $PrSb_xTe_{2-x}$ compositions measured with current flow within the *ab* plane. At zero magnetic field, all compositions show non-metallic behavior under zero magnetic field, with resistivity increases upon cooling. Providing the presence of two structure transitions with composition for $PrSb_xTe_{2-x}$ as described above, it is not surprising that the evolution of resistivity is not systematic. Notably, for low Sb compositions showing with remarkable orthorhombic structural distortion (see Fig. 1d), resistivity increases abruptly at low temperatures and reaches high values at 2 K in the $x$ = 0.12 (~80 Ω cm), 0.30 (~50 Ω cm), and 0.45 (~600 Ω cm) samples. In contrast, samples with high Sb content approaching or in the tetragonal phase region (i.e., $x > 0.6$) show enhanced metallicity as manifested by less abrupt resistivity upturns and reduced resistivity values. Those observations suggest the evolution of electronic structure with Sb content. In addition, such metallicity enhancement has also been observed in a few $Ln$Sb$_x$Te$_{2-x}$ ($Ln$ = Sm, Ce, and Gd) [60,66,70].

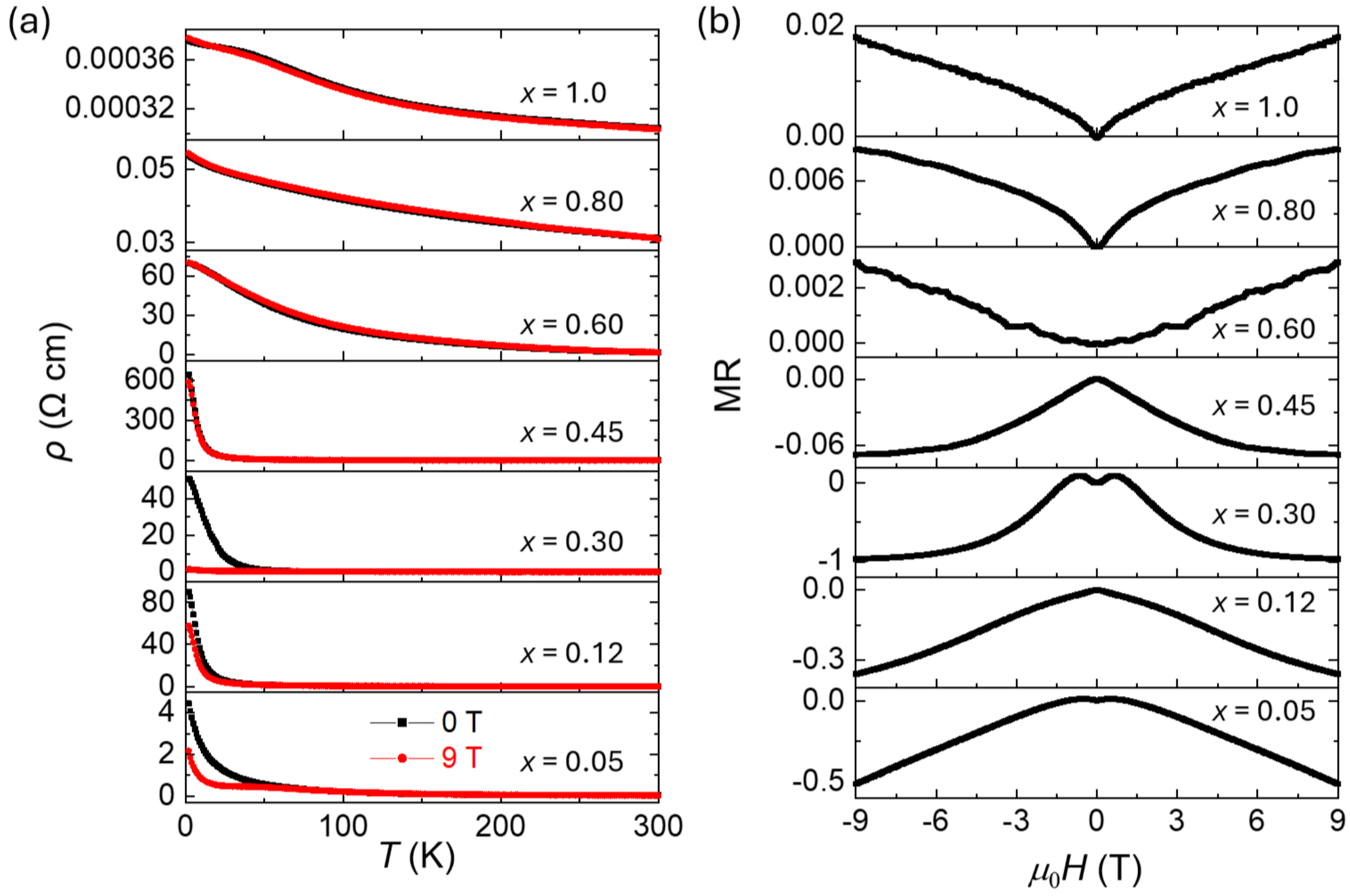

**Figure 2** (a) Temperature dependent in-plane resistivity for various $PrSb_xTe_{2-x}$ compositions from $x$ = 0.05 to $x$ = 1. (b) Normalized MR measured at $T$ = 2 K for various $PrSb_xTe_{2-x}$ compositions from $x$ = 0.05 to $x$ = 1.

The application of the magnetic field along *c*-axis modifies electronic transport. As shown in Fig. 2, samples with high Sb contents (i.e., $x$ = 0.6) display weak resistivity enhancement under field. The magnetoresistance (MR) normalized to the *zero-field resistivity value*, i.e., $|[\rho(H)-\rho(H=0)]|/\rho(H=0)$, is less than 2% at 2 K and 9 T. Unlike the positive MR for high-Sb samples, samples with less amount of Sb ($x \leqslant 0.45$) all show negative MR. Interestingly, the negative MR is rather large for $PrSb_{0.3}Te_{1.7}$, reaching 97% at 9 T. This MR value corresponds to 3,000% if normalized to the resistivity value at 9 T. To better understand the unusual magnetotransport in this composition, we have performed systematic temperature and magnetic field dependence measurements. As shown in Fig. 3a, for the $x$ = 0.3 sample, temperature dependence of resistivity measured under various magnetic fields clearly reveals resistivity suppression by magnetic field at low temperatures. The zero-field resistivity displays typical non-metallic transport with monotonic resistivity increase at low temperatures. Applying magnetic field efficiently suppresses resistivity. Though fully metallic transport behavior is not achievable up to 9 T, the strong suppression of resistivity leads to significant negative MR, which can be better seen in field-dependent resistivity measurements presented in Fig. 3b. Though the overall MR is negative, the $x$ = 0.3 sample displays a clear resistivity dip with small positive MR in the low field region (below 1.5 T), which is followed by sharp drop in the resistivity at higher fields. This resistivity dip is pronounced at 2 K and gradually weakens at elevated temperatures, becoming unobservable above 20 K. Such low field MR behavior is reminiscent of weak antilocalization phenomenon, which might occur in the presence of strong spin-orbital coupling [71–74] and thus it is expected for $PrSb_xTe_{2-x}$ given the constituted heavy elements. The strong high field MR and low field dip is reproducible in multiple samples of nearly identical composition. However, the similar low field MR dip is not present in every $PrSb_xTe_{2-x}$ composition (Fig. 2b). Therefore, further experimental and theoretical studies are needed to clarify it.

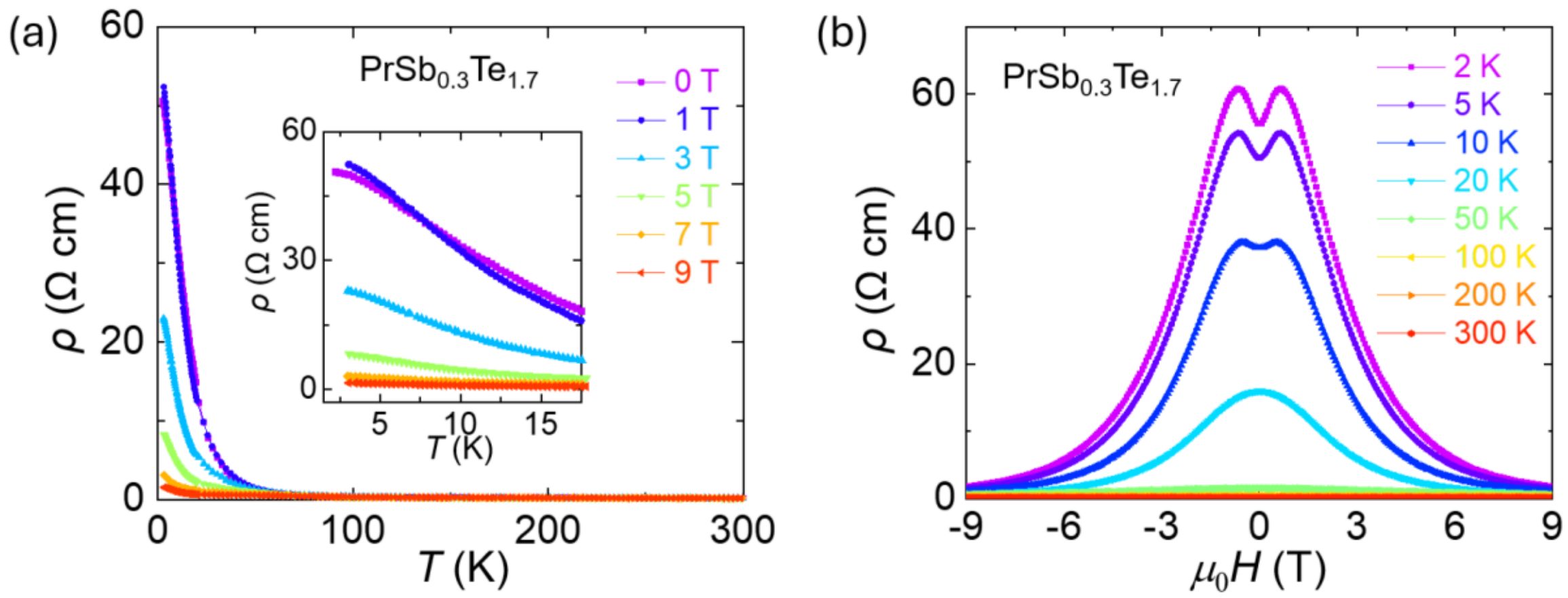

**Figure 3** (a) Temperature dependent resistivity for $PrSb_{0.3}Te_{1.7}$ measured under various magnetic fields from 0 to 9 T perpendicular to the *ab*-plane. Inset: zoom-in of the low temperature data. (b) Field dependent resistivity for $PrSb_{0.3}Te_{1.7}$ measured at various temperatures from 2 to 300 K.

To better understand the composition dependent magnetotransport and the presence of strong negative MR in the $x$ = 0.3 sample, Hall effect measurements were performed for various compositions, as shown in Fig. 4. For Sb-rich samples, such as $x$ = 0.8 (Fig. 4a) and 0.6 (Fig. 4b), the Hall resistivity $\rho_{yx}$ ($H$) displays linear field-dependence with positive slope from 2 to 300 K, indicating hole-dominant transport in these compositions. With reducing the Sb amount, owing to the inevitable mixture of the longitudinal resistivity component, it is difficult to obtain meaningful Hall effect data at low temperatures for samples with strong MR. Therefore, only data above 100

K were used for analyzing charge transport. Interestingly, reducing the Sb content in PrSbTe shifts the transport mechanism to be predominantly electron-driven particularly for the composition with Sb content $x \leqslant 0.3$ as shown in Figs. 4d and 4e. As summarized in Fig. 4f, the hole density is maximized near $x = 1$ (~$10^{24}$ $cm^{-3}$), which symmetrically reduces with lowing Sb content, and switches to electron-type transport for the $x = 0.3$ sample showing the lowest carrier density of ~$10^{22}$ $cm^{-3}$. Further decreasing Sb content enhances electron density, reaching ~$10^{25}$ $cm^{-3}$ toward the x = 0 composition. Such behavior indicates an electron-doping scenario upon increasing Te content in $PrSb_xTe_{2-x}$. Importantly, this indicates a major transformation in the electronic structure around the $x = 0.3$ composition, which is also accompanied by the most pronounced orthorhombic distortion observed in this composition. Further electronic band structure calculations and angular-resolved photoelectron emission experiments are needed to clarify the evolution of electronic structures with varying Sb content.

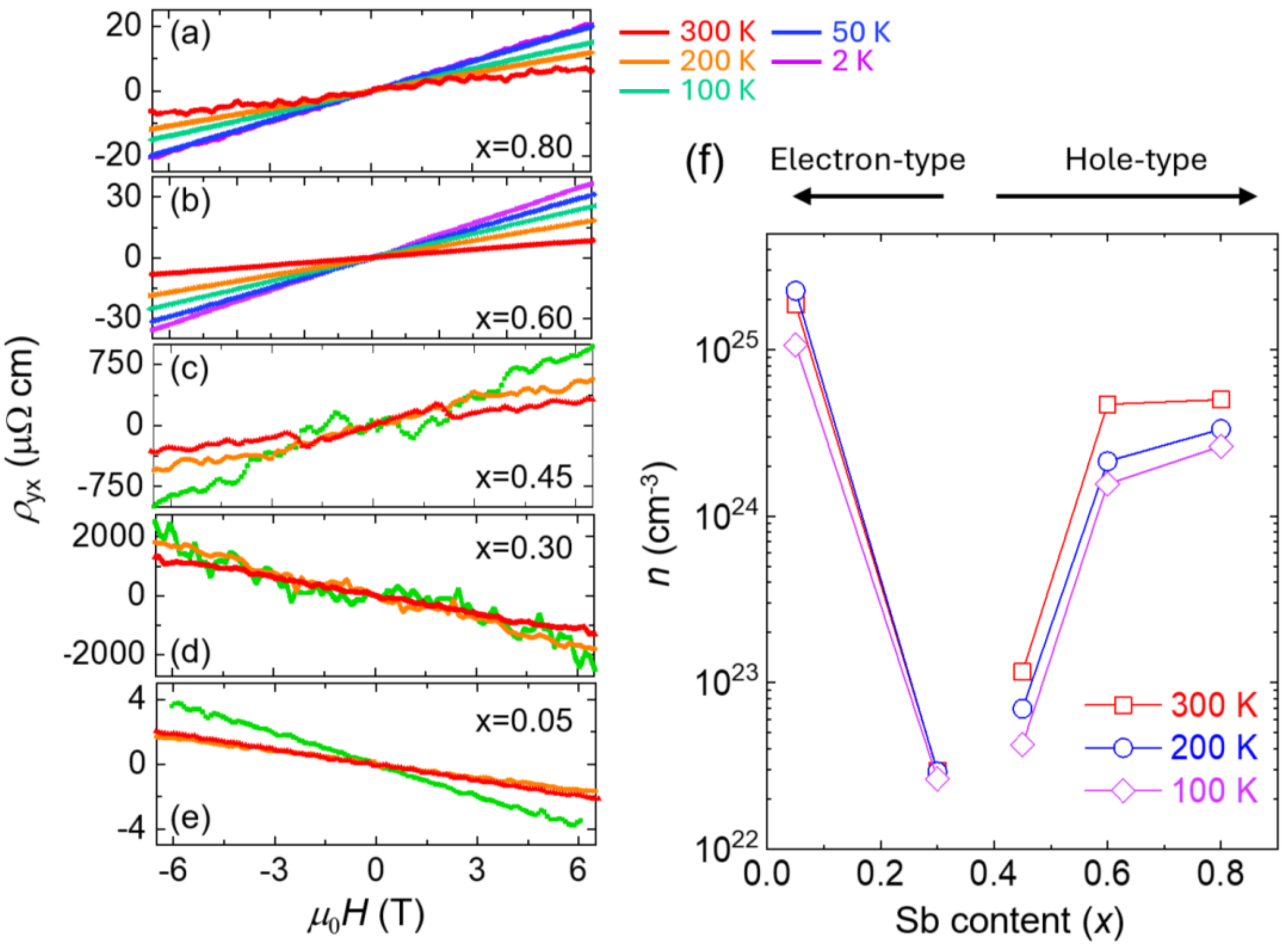

**Figure 4** (a-e) Hall effect measurements for various compositions. (f) Carrier density and type as a function of Sb content for $PrSb_xTe_{2-x}$.

In this work, we focus on strong negative MR at higher fields for the $x$ = 0.3 sample. Since it occurs for transverse MR measured with field perpendicular to the current direction, chiral anomaly induced negative MR in TSMs [10–12,18,19] should not be applicable. In fact, similar properties have also been observed in a few magnetic compounds such as $EuCd_2P_2$ [75], $EuMnSb_2$ [76,77], $EuTe_2$ [78,79], $Mn_3Si_2Te_6$ [80,81], and GdPS [47]. The negative MR in these materials has been proposed to be related to magnetism in some manners, either though magnetic fluctuations or band structure modification by magnetism or spin rotation. Therefore, we have characterized the magnetic properties for $PrSb_{0.3}Te_{1.7}$.

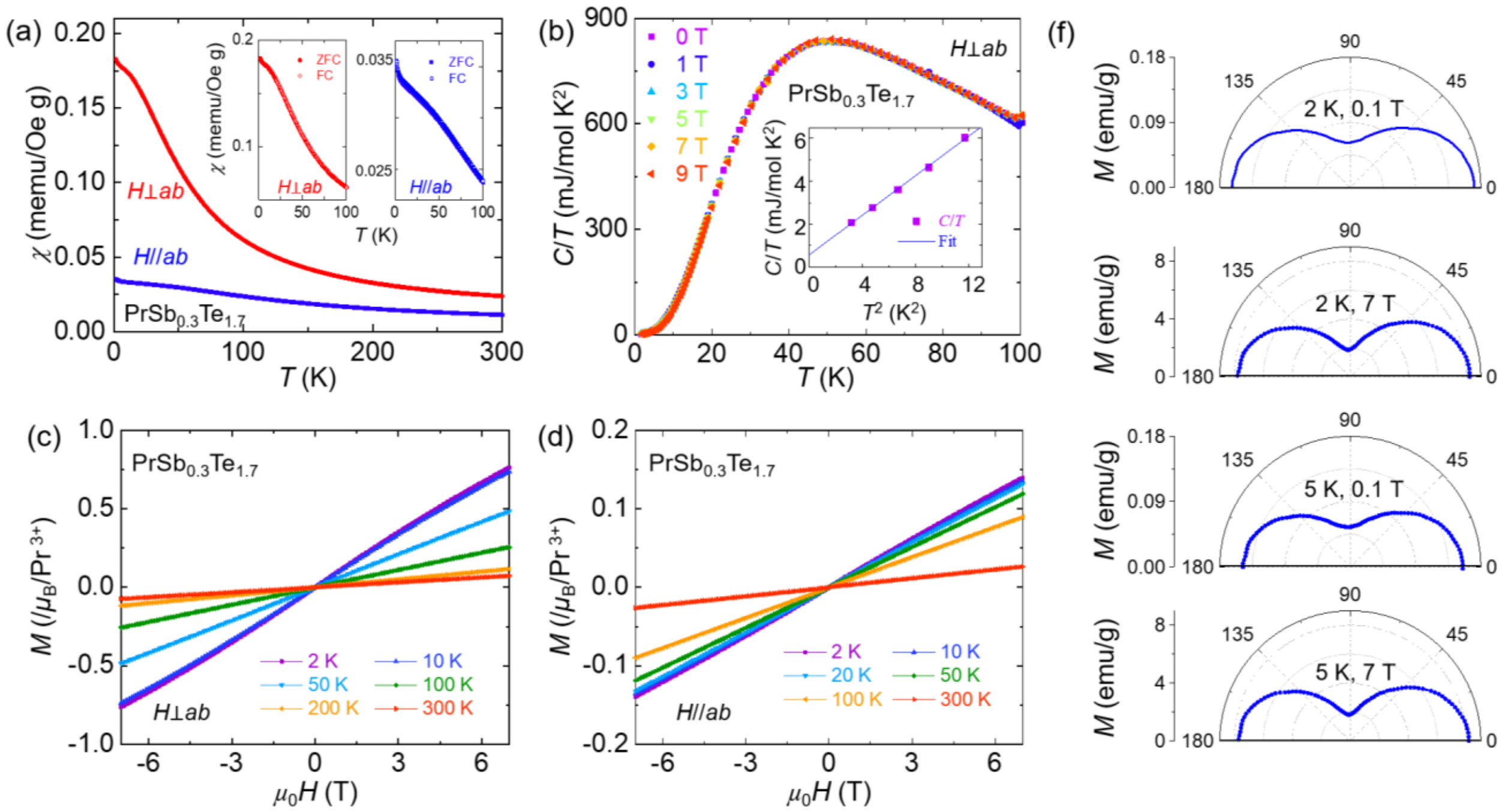

**Figure 5** (a) Magnetic susceptibility for $PrSb_{0.3}Te_{1.7}$ measured under out-of-plane ($H \perp ab$) and in-plane (*H*//*ab*) magnetic fields of 0.1 T. Inset: Zero-field-cooling (ZFC) and field-cooling (FC) magnetic susceptibility measured at 0.1 T for both magnetic field orientations. (b) Temperature dependent specific heat divided by temperature *C/T* for $PrSb_{0.3}Te_{1.7}$ under various out-of-plane fields. Inset: linear fit to the low temperature data to $C/T = \gamma + \beta T^2$. (c) and (d) Field dependent isothermal magnetization of $PrSb_{0.3}Te_{1.7}$ measured under (c) out-of-plane ($H \perp ab$) and (d) in-plane (*H//ab*) magnetic field orientations at various temperatures. (f) Angular dependent magnetization for $PrSb_{0.3}Te_{1.7}$ measured at temperatures of 2 and 5 K and fields of 0.1 and 7 T.

As shown in Fig. 5a, the temperature dependent magnetic susceptibility for $PrSb_{0.3}Te_{1.7}$ measured with both out-of-plane ($H \perp ab$-plane) and in-plane (*H*//*ab*) fields of 0.1 T exhibit monotonic increases upon cooling, without any feature of magnetic transition down to 1.8 K. The zero-field cooling (ZFC) and field-cooling (FC) susceptibilities do not reveal any irreversibility at low temperatures, as shown in the inset of Fig. 5a. Above 150 K, for both magnetic field directions, susceptibility can be fitted by the modified Curie-Weiss law $\chi_{mol} = \chi_0 + \frac{C}{T-\Theta}$, where $\chi_0$, $C$ and $\Theta$ are the temperature-independent part of the susceptibility, the Curie constant, and the Weiss temperature, respectively. From the fits, the estimated effective moments $\mu_{eff} = \sqrt{\frac{3CK_B}{N_A}}$ and Weiss temperatures $\Theta$ are 3.67 $\mu_B$ and $-65$ K for $H||ab$, and 3.54 $\mu_B$ and 20 K for $H \perp ab$. Such effective moment values are very close to the theoretical value of 3.58 $\mu_B$ for $Pr^{3+}$. The negative Weiss temperature for *H*//*ab* and the lack of irreversibility implies AFM correlations. Here Weiss temperature changes sign for $H \perp ab$, which might be caused by finite ferromagnetic correlations along the *c*-axis. The different signs for Weiss temperature for *H*//*ab* and $H \perp ab$ may occur in the presence of strong magnetic anisotropy, as has been observed in Ising-type magnetic system $FePS_3$ [82]. Indeed, susceptibilities measured with $H//ab$ and $H \perp ab$ do show significant anisotropy in the entire temperature range, as shown in Fig. 5a.

At low temperatures, the lack of well-defined magnetic transition in $PrSb_{0.3}Te_{1.7}$ above 1.8 K has also been observed in stoichiometric PrSbTe [40,41,45], which is distinct from other magnetic *Ln*SbTe [34–36,38,39,42,43,43–46,56,57]. The absence of long-rang magnetic order is further verified by heat capacity measurements. As shown in Fig. 5b, no-transition like feature can be observed from 1.8 to 100 K in heat capacity measured under various magnetic field from 0 to 9 T. Therefore, heat capacity $C$ at low temperatures can be attributed to electronic ($\gamma T$) and phonon ($\beta T^3$) contributions. Fitting the low temperature data to $C/T = \gamma + \beta T^2$ (Fig. 5b, inset) yield a Sommerfeld coefficient of $\gamma = 0.57$ mJ/mol K$^2$ and $\beta = 0.46$ mJ/mol K$^4$. Many *Ln*SbTe compounds possess large $\gamma$ of 41 – 382 mJ/mol K$^2$ for various $Ln$ = Ce [56], Nd [34], Sm [39,60], and Ho [38]. GdSbTe displays a relatively small $\gamma$ of 7.6 mJ/mol K$^2$ [62], which is still an order of magnitude larger than our $PrSb_{0.3}Te_{1.7}$. Similarly, stoichiometric PrSbTe also displays small $\gamma$ value of 7.45 [40] or 2.62 [41] mJ/mol K$^2$ in different studies. Also, small $\gamma$ below 1 mJ/mol K$^2$ is seen in DySbTe [44] and TbSbTe [42], as well as non-magnetic LaSbTe [34,83,84]. From the obtained $\beta$ value, the Debye temperature can be estimated by $\Theta_D = [12\pi^4 nR/(5\beta)]^{1/3} \sim 233$ K, which is slightly higher than that of the stoichiometric PrSbTe [40,41].

As shown in Fig. 5b, applying magnetic field does not notably change heat capacity for $PrSb_{0.3}Te_{1.7}$. Similarly, the isothermal field-dependent magnetization measured with both $H \perp ab$ (Fig. 4c) and $H//ab$ (Fig. 5d) displays linear field dependence without any clear feature for metamagnetic transitions. This is also observed for the stoichiometric PrSbTe [59] but distinct from other *Ln*SbTe compounds with well-defined AFM order above 2 K and metamagnetic transitions such as GdSbTe, NdSbTe, CeSbTe, HoSbTe, SmSbTe, TbSbTe, and DySbTe [34,38,39,42,44,56,62]. Such linear magnetization is not surprising since a magnetically ordered state is not achieved above 1.8 K. On the other hand, magnetization at 2 K and 7 T reaches 0.75 $\mu_B$ per $Pr^{3+}$ under $H \perp ab$ (Fig. 5c). Such a substantial value is consistent with the strong susceptibility upturn at low temperatures (Fig. 5a), implying strong magnetic correlations. In contrast, magnetization measured with $H//ab$ is much smaller (0.14 $\mu_B$ per $Pr^{3+}$ at 2 K and 7 T), indicating significant magnetic anisotropy. To better clarify the magnetic anisotropy in $PrSb_{0.3}Te_{1.7}$, we measured angular dependence for magnetization with the magnetic field rotate from the out-of-plane direction ($H \perp ab$, defined as 0 degree) to in-plane ($H//ab$, defined as 90 degree). As shown in Fig. 5f, from 0.1 T to 7 T (the maximum field of our instrument), magnetization maximizes for $H \perp ab$ while monotonically reduces when rotating field toward in-plane direction,

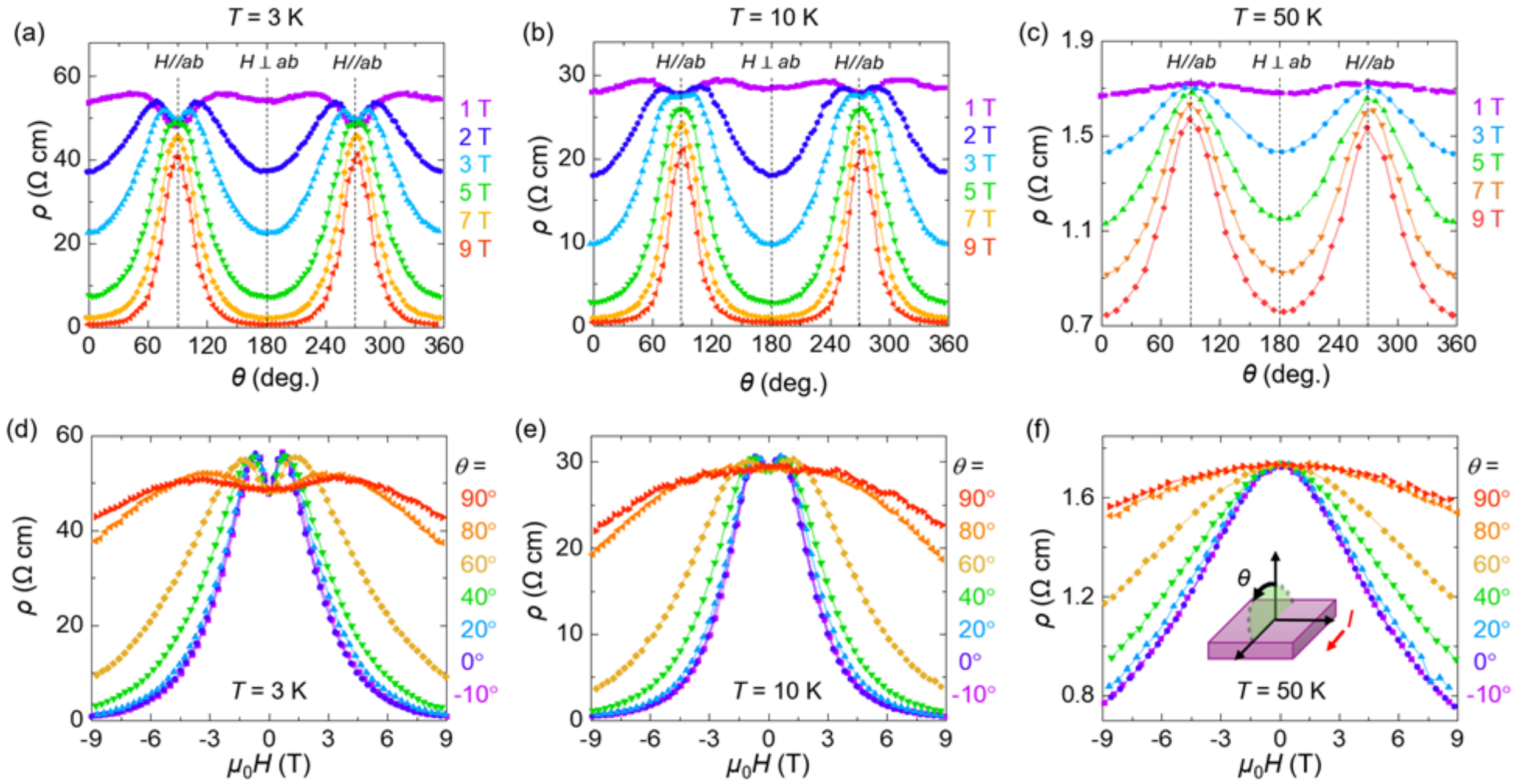

leading to a two-fold anisotropy.

**Figure 6** (a-c) Angular dependence of in-plane resistivity $\rho$ at $T$ = 3 K (a), 10 K (b) and 50 K (c) measured under various magnetic fields from 1 to 9 T. (d-f) Field dependance of resistivity at $T$ = 3 K (d), 10 K (e) and 50 K (f) measured with various fixed magnetic field orientations. Inset in (f): the measurement setup.

With the understanding of the AFM correlations in $PrSb_{0.3}Te_{1.7}$, next we try to examine the interplay between magnetism and transport. Owing to the strong anisotropy in magnetism as stated above, we have measured magnetotransport with rotating the magnetic field orientation. Figures 6a-c displays angular dependent MR measured at various fixed magnetic fields from 1 to 9 T. The schematic of field orientation rotation is shown in the inset of Fig. 6f. As shown in Fig. 6a, the MR in $PrSb_{0.3}Te_{1.7}$ generally displays a two-fold anisotropy. At 3 K and 9 T, resistivity is maximized (41 Ω cm) for $H//ab$ ($\theta$ = 90°), and quickly drops when magnetic field is deviated from this in-plane direction, reaching minimum (0.7 Ω cm) when field is perpendicular to the *ab*-plane ($\theta$ = 0°). Such behavior indicates that the perpendicular field component plays a key role in strong resistivity suppression. The relative change of the angular MR, $[\rho(H//ab)-\rho(H\perp ab)]/\rho(H\perp ab)$, reaches ~5,800%. Such two-fold anisotropy in magnetotransport is reminiscent of the magnetism anisotropy described above (Fig. 5f), implying intimate correlations between magnetism and transport in $PrSb_{0.3}Te_{1.7}$.

The two-fold anisotropy with maximum negative MR occurring for $H\perp ab$ persists with lowering the magnetic field to 5 T. Below 3 T, a slight positive MR occurs, leading to a dip at $H//ab$, as shown in Fig. 6a. This is consistent with the low-field positive MR in field dependent measurements shown above (Fig. 3b), and can be better illustrated by MR measurements at fixed angles. As shown in Fig. 6d, at $T = 3$ K, the low field resistivity dip and the corresponding positive MR is the most significant under a perpendicular field configuration ($\theta = 0°$), and becomes gradually broadened with rotating field toward the *ab*-plane. Such low field MR dip is suppressed with increasing temperature, becoming rather weak at 10 K and not observable at 50 K, as shown in Figs. 6b,c,e,f. In contrast, the substantial negative MR and the two-fold anisotropy persists against heating.

The coincidence between magnetism and transport anisotropies implies the interplay between them. In magnetic materials, the strong negative MR might arise from a spin valve effect [1], suppression of spin scattering, magnetic polaron [78,85–87], or modification of electronic structures under magnetic field. Providing that the negative MR is probed in in-plane transport and the Sb layer contributes significantly to electronic transport in *Ln*SbTe materials [39,66], a spin valve mechanism can be ruled out. For spin scattering, it is possible providing the observed strong magnetic correlations. However, this mechanism cannot explain the composition sensitive MR in $PrSb_xTe_{2-x}$ as shown in Fig. 2b. Specifically, the $x = 0.3$ sample and the stoichiometric PrSbTe [40] display similar magnetization, but show distinct negative and positive MR (Fig. 2b), respectively. Therefore, considering the structure evolution of $PrSb_xTe_{2-x}$, the modification of electronic structure by magnetic field is possible. This mechanism has already been reported in a few compounds showing strong negative MR, such as $EuMnSb_2$ [77], $EuTe_2$ [78,79], and $Mn_3Si_2Te_6$ [80,81], as well as GdPS which is structurally similar to $PrSb_xTe_{2-x}$ [47]. In these materials, strong magnetization couples with electronic band structure through spin orbital coupling or exchange splitting, which reduces or closes the band gap to enhance electronic conduction and even results in an insulator-to-metal transition. In other well-studied $LnSb_xTe_{2-x}$ materials, charge density wave has been identified in orthorhombic off-stoichiometric compositions [63,65,70], which might open a small or partial gap [69]. Therefore, a similar gap reduction scenario might be applicable to $PrSb_xTe_{2-x}$, which is consistent with the similar anisotropies of MR and magnetism. Nevertheless, the strong composition-dependent MR behavior (Fig. 2) indicates that magnetism along cannot fully explain the strong MR in $PrSb_{0.3}Te_{1.7}$. It is worth noting that this composition also displays the strongest orthorhombic distortion, as manifested by the largest difference between the in-plane lattice constants *a* and *b* (Fig. 1d). This might lead to the strongest modification of the electronic structure, especially considering the possible charge density wave that might gap the electronic structure. Strong negative MR has also been observed in another *Ln*SbTe compound $CeSb_{0.11}Te_{1.90}$, which has been suggested to correlate to charge density wave [70]. Though the nature of such correlation remains unclear, it appears to be generic to $LnSb_xTe_{1-x}$ materials. Additionally, in stoichiometric *Ln*SbTe, the nodal-line Dirac crossing near the Fermi energy in *Ln*SbTe compounds are protected by the mirror symmetry of the *P*4/*nmm* tetragonal lattice [49]. The symmetry reduction to orthorhombic lattice, however, does not affect the mirror planes along crystallographic *a*- and *b*- axes (see Fig. 1c). Hence it is not clear how orthorhombic distortion modifies the electronic structure. Nevertheless, the change of carrier type and the minimum carrier density around the $x = 0.3$ composition revealed by Hall effect does indicate strong band structure modifications for this composition. Therefore, it is likely that the magnetism and structure distortion together gives rise to the strong negative MR in $PrSb_{0.3}Te_{1.7}$. Additional theoretical and experimental studies to clarify the impact of composition and lattice structure, as well as the impact of magnetism and exchange splitting are needed to fully clarify the origin of the strong negative MR in the $x = 0.3$ sample and its composition dependence. Additionally, considering the substitution of Sb plane by Te in off-stoichiometric $PrSb_xTe_{2-x}$, the impact of atomic ordering, whether the formation of supercell or chemical short-range orders, needs to be further clarified as such structure orderings can significantly affect the electronic structure. Particularly, the chemical short-range orders, which occur in doped systems and characterizes the deviation of the atomic distribution from the perfect random distributions, can significantly affect the band gaps in semiconductors [88,89].

In summary, we have successfully grown $PrSb_xTe_{2-x}$ and characterized its structural and transport properties, from which we have identified the strong negative MR in the $x = 0.3$ composition. The careful examination of magnetism and magnetotransport reveals similar anisotropies and suggests the correlation between them. Such negative MR possibly arises from the band structure modification under magnetic field, with further engineering the materials to enhance the temperature and reduce magnetic field to achieve large MR, this material can provide a platform to explore the implementation of topological materials in spintronic applications.

## Acknowledgments

Crystal growth and magnetotransport were supported by the U.S. Department of Energy, Office of Science, Basic Energy Sciences program under grant No. DE-SC0022006. We acknowledge the MonArk NSF Quantum Foundry for magnetic property measurements using MPMS3 SQUID, which is supported by the National Science Foundation Q-AMASE-i program under NSF award No. DMR-1906383. R.B. acknowledge μ-ATOMS, an Energy Frontier Research Center funded by DOE, Office of Science, Basic Energy Sciences, under Award No. DE-SC0023412 for part of the magnetic property analysis. J. S. acknowledges the support from NIH under award P20GM103429 for the powder XRD experiment. J. W. acknowledges DMR2316811 for single crystal x-ray refinement.